\documentclass[10pt]{article}
\usepackage[cp1251]{inputenc}
\usepackage[english]{babel}
\usepackage{graphicx}
\usepackage{amssymb}
\usepackage{amsmath}
\usepackage{amsthm}
\usepackage{amsfonts}
\usepackage{amssymb}
\title{{\bf \Large  Cosmic Evolution in Fractional Action Cosmology}\\
{\normalsize ~~{\bf V.\,K. Shchigolev}\thanks{E-mail:
vkshch@yahoo.com}}\\
{\small {\it Ulyanovsk State University, 42 L. Tolstoy Str.,
Ulyanovsk 432000, Russia}}\\
\vspace{2mm}
\small \begin{quote}{\bf Abstract} --  For the fractional  action cosmological model, derived earlier by the author from the variational principle for a fractional action functional, the exact solutions  are obtained. The case of a quasi - vacuum state of matter that fills the universe is considered. Moreover, on the basis of specific {\it ansatz} proposed in this paper for the cosmological term,  the class of exact solutions  of  the model equations  is obtained. Examples for some given laws of the cosmological term evolution  are provided. Besides,  a formula for the effective equation of state is derived,  and  the deceleration parameter of the obtained models is studied.\\
\vspace{2,5mm}
{\bf PACS numbers}: 98.80.-k; 98.80.Jk; 04.20.Jb.\\
{\bf Key words}: Cosmological Models, Fractional Action, Exact Solutions, Accelerated Expansion.\\
\end{quote}}
\date{}

\begin{document}
\maketitle
\vspace{-20mm}
\section{Introduction}

As well known, the fractional action cosmology (FAC) is the new field of research based on the principles and formalism of the fractional calculus applied to cosmology. In this paper, we present the results of study of some exact models in FAC constructed by the author on the basis of the fractional action functional in \cite{Shchigolev}. These models are followed from  the fractional variational principle for the dynamical field theories in general and the theory of gravity formulated by El - Nabulsi (see, e.g., \cite{1Nabulsi}, \cite{2Nabulsi} and references therein). In this approach, the action integral $S_L [q]$ for the Lagrangian $L(\tau, q(\tau), \dot q(\tau))$ is written as a fractional integral \cite{Uchaikin}:
\begin{equation}
\label{1} S_L [q_i]=\frac {1}{\Gamma
(\alpha)}\int\limits_{t_0}^{t} L(\tau,q_i (\tau),\dot
q_i(\tau))(t-\tau)^{\alpha-1} d\tau ,
\end{equation}
which is essentially  the Riemann-Stieltjes  integral for the function $L$ at fixed $t$ with an integrating function ${\displaystyle g_t(\tau) = \frac{1}{\Gamma (1 + \alpha)} [t^{\alpha} - (t-\tau)^{\alpha}]}$. This function has the following scaling property: $g_{\mu t}(\mu \tau) = \mu^{\alpha} g_t(\tau), ~ ~ \mu > 0$. It is worth to note that recently Calcagni \cite{1Calcagni}, \cite {2Calcagni} gave a quantum gravity in a fractal universe and then investigated cosmology in that framework. That theory is Lorentz invariant, power-counting renormalizable and free from ultraviolet divergence. The action in this model is also a Lorentz-covariant and is equipped with a Stieltjes measure, but the model equations are somewhat different from the basic equations of FAC. The holographic, new agegraphic and ghost dark energy models in the framework of
fractal cosmology are investigated in \cite{Karami}.   The fractal universe in which dark energy interacting with dark matter is considered in \cite{Yerokhin}.

In the present work, we aim to  develop further the FAC model proposed by the author in \cite{Shchigolev} and  obtain the new exact solutions to this model.
Exact solutions of the modified Friedmann equations in FAC are important to analyze the possible  behavior of these models in aspect of the accelerated cosmological expansion in aspect of the recent observational data. To the best of our knowledge, until now, a small number of such solutions is known. Moreover,  these solutions are mainly based on a predetermined evolution law of the scale factor \cite{Debnath}, \cite{Jamil},  although it would be logical to start with some reasonable physical or mathematical  proposals, and only then  to analyze the behavior of the scale factor. In this paper, we obtain several exact solutions from the assumption of a vacuum-like state of matter or from a certain,  rather general, relation between the cosmological term and the Hubble parameter, which   were previously used   by many authors in some particular cases. Besides, we derive a formula for the effective equation of state (EoS) in FAC. Graphical illustrations of all obtained models are also given. Our models demonstrate the new types of cosmic evolution, which are not standard for the cosmological models,  based on General Relativity.

\section{The model equations of FAC}

Following the definition (\ref{1}), the modified fractional effective action in a spatially flat Friedmann-Robertson-Walker metric, $\displaystyle ds^2 = N(t)^2 dt^2 - a^2(t) \delta_ {ik} dx^i dx^k$, where $N$ is the lapse function and $a(t)$ is a scale factor, is represented by a fractional integral as follows \cite{Shchigolev}:
\begin{equation}
\label{2} S^{\alpha}_{eff}=\frac {1}{\Gamma
(\alpha)}\int\limits_{0}^{t} N \left [ \frac{3}{8\pi G}
\left(\frac{a^2\ddot a}{N^2}+\frac{a\dot a^2}{N^2}-\frac{a^2\dot a
\dot N}{N^3}-\frac{\Lambda a^3}{3}\right)+
a^3 {\cal L}_m\right
](t-\tau)^{\alpha-1} d\tau~,
\end{equation}
where $\alpha \in (0,1)$, the over-dot denotes differentiation with respect to time $t$, and ${\cal L}_m$ is the Lagrangian density of matter characterized by the energy density $\rho$ and pressure $p$. The variation of the action (\ref{2}), followed by the choice of gauge $N = 1$, yields the modified equation of continuity,
\begin{equation}
\dot \rho+3\left(H+\frac{1-\alpha}{3 t}\right)(\rho+p)=0,\label{3}
\end{equation}
and the set of Euler-Poisson equations, which can take the form of the following modified Friedmann equations \cite{Shchigolev}:
\begin{eqnarray}
\rho &=& 3 H^2+3\frac{(1-\alpha)}{t}H-\Lambda{~,}\label{4}\\
p &=& -2\dot H- 3 H^2-2\frac{(1-\alpha)}{t}H -
\frac{(1-\alpha)(2-\alpha)}{t^2}+\Lambda{~,}\label{5}
\end{eqnarray}
where the gravitational constant $8 \pi G = 1$, and $H(t) = \dot a / a $ is the Hubble parameter.
It is known that in standard cosmology the continuity equation for a perfect  fluid, that is the energy conservation law  for matter, follows from the Bianchi identity for the Riemann curvature tensor. In other words, in the standard cosmology where $\alpha = 1$, the continuity equation (\ref{3}) is a differential consequence of the field equations (\ref{4}), (\ref {5}). It is easy to prove that these equations also yield the modified continuity equation (\ref{3}) in the case $\alpha \ ne 1 $ , but only if the following equation is valid \cite{Shchigolev}:
\begin{equation}\label{6}
\dot H + 3 H^2  - \frac{2(4-\alpha)}{t}H
-\frac{(1-\alpha)(2-\alpha)}{t^2}
= \frac{t \dot \Lambda}{1-\alpha}.
\end{equation}
Note that equation (\ref{6}) is written down after dividing by a nonzero factor $(1 - \alpha)$, which is originally appeared on the left hand side of this equation. Therefore, in the limit of the standard cosmology of General Relativity, i.e. for $\alpha = 1$, the original equation (\ref{6}) uniquely leads to a constant cosmological term $\Lambda$, and the set of equations (\ref{4}), (\ref{5}) takes its usual form of the Friedmann equations.

The continuity equation (\ref{6}) is explicitly integrable for a perfect  fluid with the EoS  $p = w \rho$ with $w=constant$, which gives:
\begin{equation}
\label {7} \rho =
\frac{\rho_0}{a^{\displaystyle3(1+w)}t^{\displaystyle(1+w)(1-\alpha)}}~.
\end{equation}
If the constancy of the barotropic index $w$  is not required as a condition, that is supported by the recent observational data and theoretical studies , then it is more convenient to deal with the set of equations  (\ref{4}) - (\ref{6}) as the main system of dynamical equations for our model.

Then according to the expressions (\ref{4}) and (\ref{5}), the effective EoS $w_{eff} = p_{eff} / \rho_{eff} $, where $p_{eff} = p-\Lambda$ and $\rho_{eff} = \rho + \Lambda $, can be reduced to the following form:
\begin{equation}\label{8}
w_{eff}= -1-\frac{2}{3}\,\frac{\displaystyle \frac{\dot H}{H^2}-\frac{1-\alpha}{2(tH)}+\frac{(1-\alpha)(2-\alpha)}{2(tH)^2}}{\displaystyle 1+\frac{1-\alpha}{(tH)}}.
\end{equation}
It is easy to note that $w_{eff}$ coincides with the known standard expression  $w_{eff} = \displaystyle -1 - \frac{2}{3} \, \frac{\dot H}{H^2}$ in the limit $\alpha \to 1$, but can significantly differ from it in the case of fractional order of the effective action (\ref{2}). Moreover, the possibility of vanishing of the numerator in (\ref{8}) at some instant means that the model admits crossing of the so-called phantom divide $w_{eff} = -1$.  Of course, the latter does not mean that the EoS of matter is necessarily  $w = -1$  at the moment of crossing the phantom divide. It should be noted also that the effective EoS is a dynamical characteristic of the model and  gains a new definition represented by the formula (\ref{8}), but the deceleration parameter,
\begin{equation}\label{9}
q = -\frac{a^2\, \ddot a}{\dot a^2} = -1-\frac{\dot H}{H^2},
\end{equation}
is defined just as in the standard cosmology, being a kinematical parameter  of the model \cite{Starobinsky}.

\section{The quasi-vacuum EoS of matter: $w=-1$}

Let us consider  a simple example of exact solution for our model which conforms to the following condition on the EoS of matter: $w=-1$. It follows from (\ref{7}) that $\rho (t) = \rho_0 = constant$ and $p = - \rho = - \rho_0$, exactly  as in the standard cosmology of GR. Then the rest independent equations of the system (\ref{4}) - (\ref{6}) for the Hubble parameter $H(t)$ and
cosmological term $\Lambda (t)$ can be rewritten as follows:
\begin{eqnarray}
\dot H - \frac{1-\alpha}{2 t}H +\frac{(1-\alpha)(2-\alpha)}{2t^2}= 0{~,}\label{10}\\
H^2+\frac{1-\alpha}{t} H =\frac{1}{3}\rho_0+\frac{\Lambda}{3}{~.}\label{11}
\end{eqnarray}

Equation (\ref{10})  can be easily solved, and this solution for the Hubble parameter shows that it varies with time as follows:
\begin{equation}
\label{12} H = \frac{C_{\alpha}}{t}+ H_0\,
t^{\displaystyle\frac{1-\alpha}{2}}~,
\end{equation}
where
$C_{\alpha}=\displaystyle\frac{(1-\alpha)(2-\alpha)}{(3-\alpha)}\,$, and
$H_0$ is a positive constant of integration. The latter yields the scale factor as a function of time $t$:
\begin{equation}
\label{13} a = a_0\,\, t^{\displaystyle C_{\alpha}}\exp
\left(\frac{3-\alpha}{2}H_0 t^{\displaystyle \frac{3-\alpha}{2}}
\right) ~.
\end{equation}
As an illustration, the graphs of functions $H(t)$ and $a(t)$ for $\alpha = 0.8$ are shown in Fig. 1.

The effective cosmological term $\Lambda_{eff} = \Lambda + \rho_0$ as a function of time can be found from the equation (\ref{11}) in the form
\begin{equation}
\label{14} \Lambda_{eff} = 3 H_0^2 t^{\displaystyle 1-\alpha}+3H_0
\frac{(1-\alpha)(7-3\alpha)}{(3-\alpha)}\phantom{.}t^{\displaystyle-\frac{(1+\alpha)}{2}}+
\frac{3(1-\alpha)^2(2-\alpha)(5-2\alpha)}{(3-\alpha)^2}\frac{1}{t^2}.
\end{equation}
It is seen that  in the limit $\alpha \to 1$,  which  leads to  the standard cosmology, our solution (\ref{12}) - (\ref{14}) tends to the usual exponential expansion of the universe: $a = a_0 e^{\displaystyle H_0 t},~H=H_0,~\Lambda_{eff}=3 H_0^2$.

An interesting feature of the obtained solution is that according to the formula (\ref{8}) and equation (\ref{10}) the effective equation of state coincides with the equation of state of matter that is equal to $w_{eff} = -1$, but the cosmological term evolves according to the formula (\ref{14}). Even more significant difference of this model from the standard $\Lambda$CDM model is the behavior of the deceleration parameter. From the equations (\ref{9}) and (\ref{10}), it follows that
$$
q(t) = -1-\frac{1-\alpha}{2(tH)}+\frac{(1-\alpha)(2-\alpha)}{2(tH)^2}.
$$
\begin{figure}[t]
\includegraphics[width=0.47\textwidth]{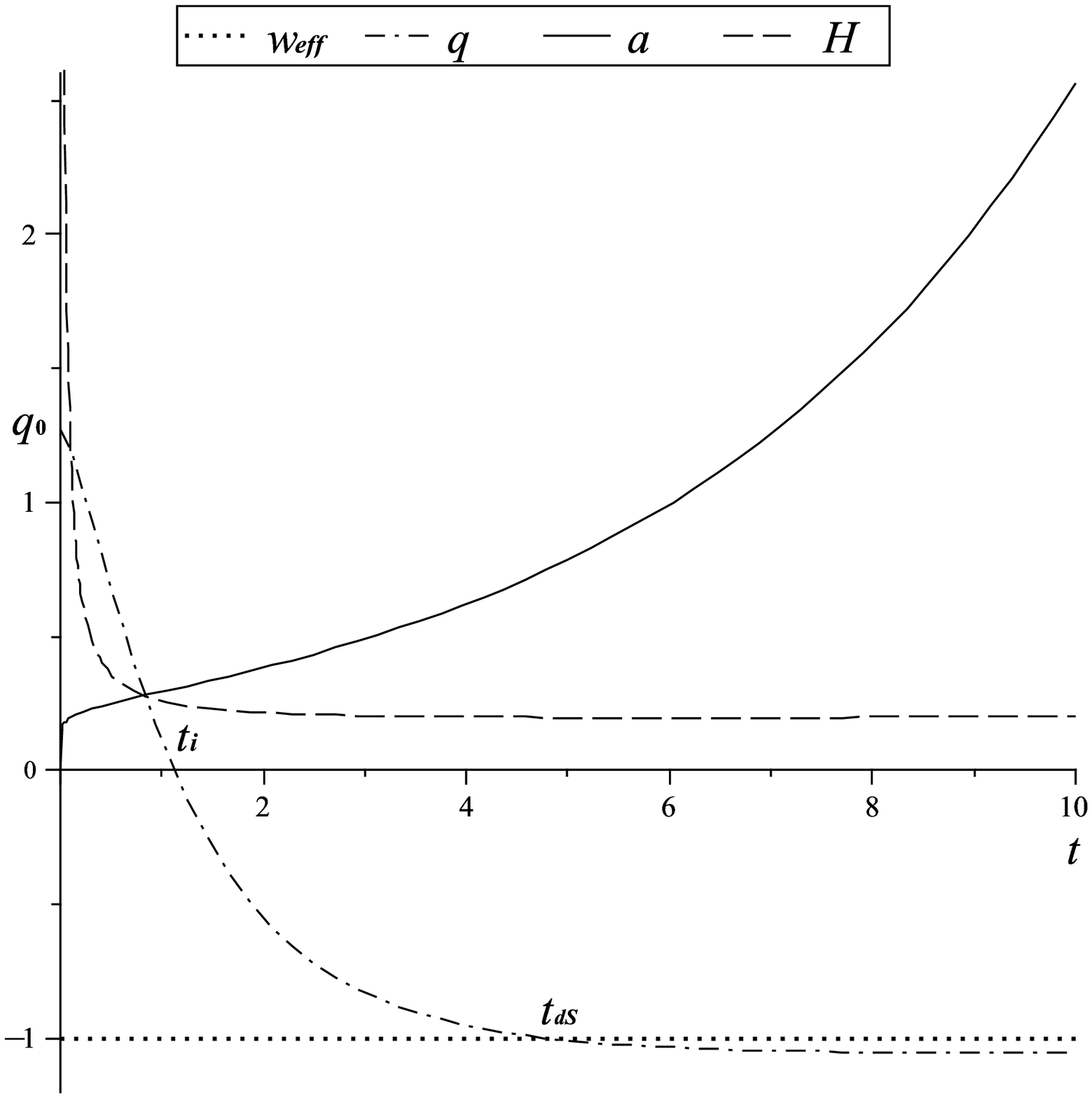} \hfill
\includegraphics[width=0.47\textwidth]{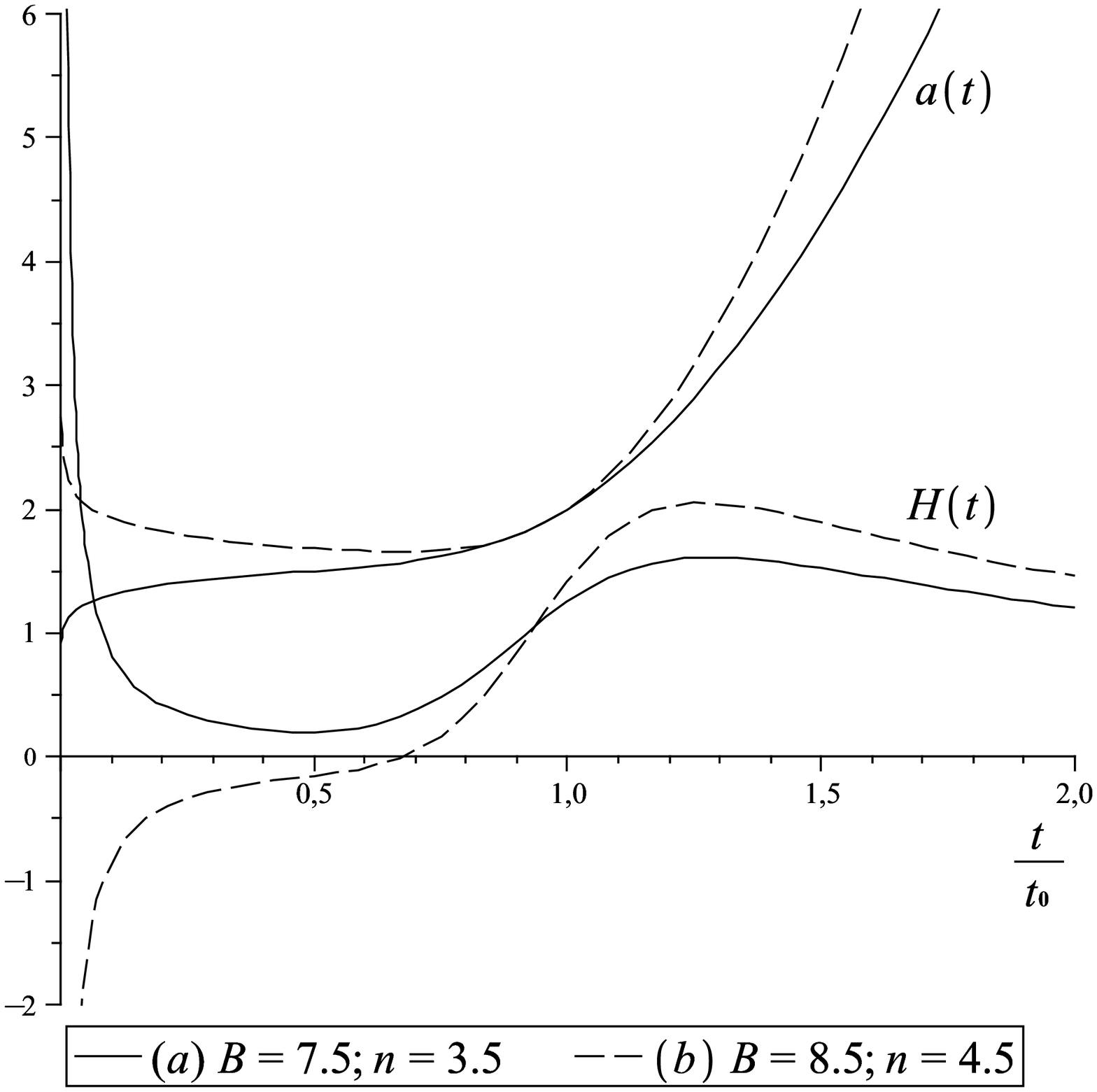}
\parbox[t]{0.47\textwidth}{\caption{The evolution of the effective equation of state $w_{eff}$, the deceleration parameter $q$, the scale factor $a$ and the Hubble parameter $H$ in the model $w = -1$. Here $\alpha = 0.8$, $t_0 = 0.25$ and $H_0 = 0.15$.}\label{fig1}} \hfill
\parbox[t]{0.47\textwidth}{\caption{The evolution of the Hubble parameter (\ref{23}) and the scale factor (\ref{24}) for different values of $B$ and $n$. In both cases, $A = 1, \, C = 3$ and $a_0 = 2 $.}\label{fig2}}
\end{figure}
Using the explicit form of expression $tH = C_{\alpha} + H_0 \, t^{(3 - \alpha)/2}$ from (\ref{12}), it is easy to show that the deceleration parameter starts to decrease from the value $ q_0 = q(t = 0) = -1 + C^{-1}_{\alpha}> 0$ down to zero,  reaching it  at some instant $ t_i $, which is easy to find from the equation $q(t_i) = 0$ as
$$
t_i= (4H_0)^{\displaystyle-\frac{2}{3-\alpha}}\Big[\sqrt{(1-\alpha)(17-9\alpha)}-\frac{(1-\alpha)(11-5\alpha)}{3-\alpha}\Big]^{\displaystyle \frac{2}{3-\alpha}}.
$$
After that, initially slow expansion is replaced by an accelerated expansion for all $t>t_i$. At time $t_{dS} = [2 (2 - \alpha)/ H_0 (3 - \alpha)]^{2 / (3 - \alpha)} $  the deceleration parameter crosses the de Sitter line $q = -1$ and then reaches its minimum  $q_{min} = -1 - [(1 - \alpha)/8(2 - \alpha)] $ at time $t_{min} = [(2 - \alpha)(5 - \alpha)/H_0(3 - \alpha)]^{2 / (3 - \alpha)}> t_{dS}$. From this instant, the deceleration parameter asymptotically tends to $ q_{\,\infty} = -1$, staying in the domain $q <-1$.   The plots of $q$ and $w_{eff}$ versus time are shown in Fig. 1.

For the standard $\Lambda$CDM model, i.e. when $\alpha = 1$, we find $t_i = 0$, which corresponds to the accelerated expansion  at any time $t \ge 0$,  accompanying by the constant  deceleration parameter $q = -1$.

\section{Models with a given $\Lambda(t)$}

Let us obtain a class of exact solutions of the equation (\ref{6}), assuming that  the cosmological term $\Lambda(t)$  is given by some condition, and it is a known function of time.  Then we make the following substitution in the equation (\ref{6}):
\begin{equation}\label{15}
x=\ln (t/t_0) \Leftrightarrow t=t_0 \exp(x);\,\,\, Y(t) = t\, H(t),
\end{equation}
where $t_0> 0$ is a constant. As a result, it can be rewritten as
\begin{equation}\label{16}
Y'(x) - (9-2\alpha) Y(x) + 3 Y^2(x) -(1-\alpha)(2-\alpha)=\frac{t_0^2}{1-\alpha}e^{\displaystyle 2x}\Lambda '(x),
\end{equation}
where the prime denotes the derivative with respect to $x$. From the form of this equation, we can assume that there exists a class of solutions for this model, for which the cosmological term satisfies the following equation:
\begin{equation}\label{17}
\Lambda '(x) = \frac{1-\alpha}{t_0^2} e^{\displaystyle -2x}\Big(k_1 Y'(x) + k_2 Y(x)+k_3 Y^2(x) +k_4\Big),
\end{equation}
where $ k_i $ are arbitrary constants. Indeed, substitution of (\ref{17}) in the equation (\ref{16}) leads the latter to the form:
\begin{equation}\label{18}
A Y'(x) - B Y(x) + C Y^2(x) = D,
\end{equation}
where the coefficients are:
\begin{equation}\label{19}
A = 1-k_1,\,B = 9-2\alpha +k_2,\,C = 3-k_3,\,D = (1-\alpha)(2-\alpha) + k_4.
\end{equation}
Therefore, all models in which the cosmological term would be obtained from the {\it ansatz} (\ref{17}) have the same type of solution and evolve similarly. The general solution of (\ref{18}) can be written as
\begin{equation}\label{20}
Y(x) = \frac{1}{2C}\left[B+\sqrt{B^2+4CD}\,\tanh \left(\frac{\sqrt{B^2+4CD}}{2A}x\right)\right]
\end{equation}
where $B^2 + 4 CD> 0$ and $C \ne 0$, and the constant of integration is absorbed by the arbitrary constant $t_0$, which can always be done in view of the definition (\ref{15}).
\begin{figure}[t]
\includegraphics[width=0.47\textwidth]{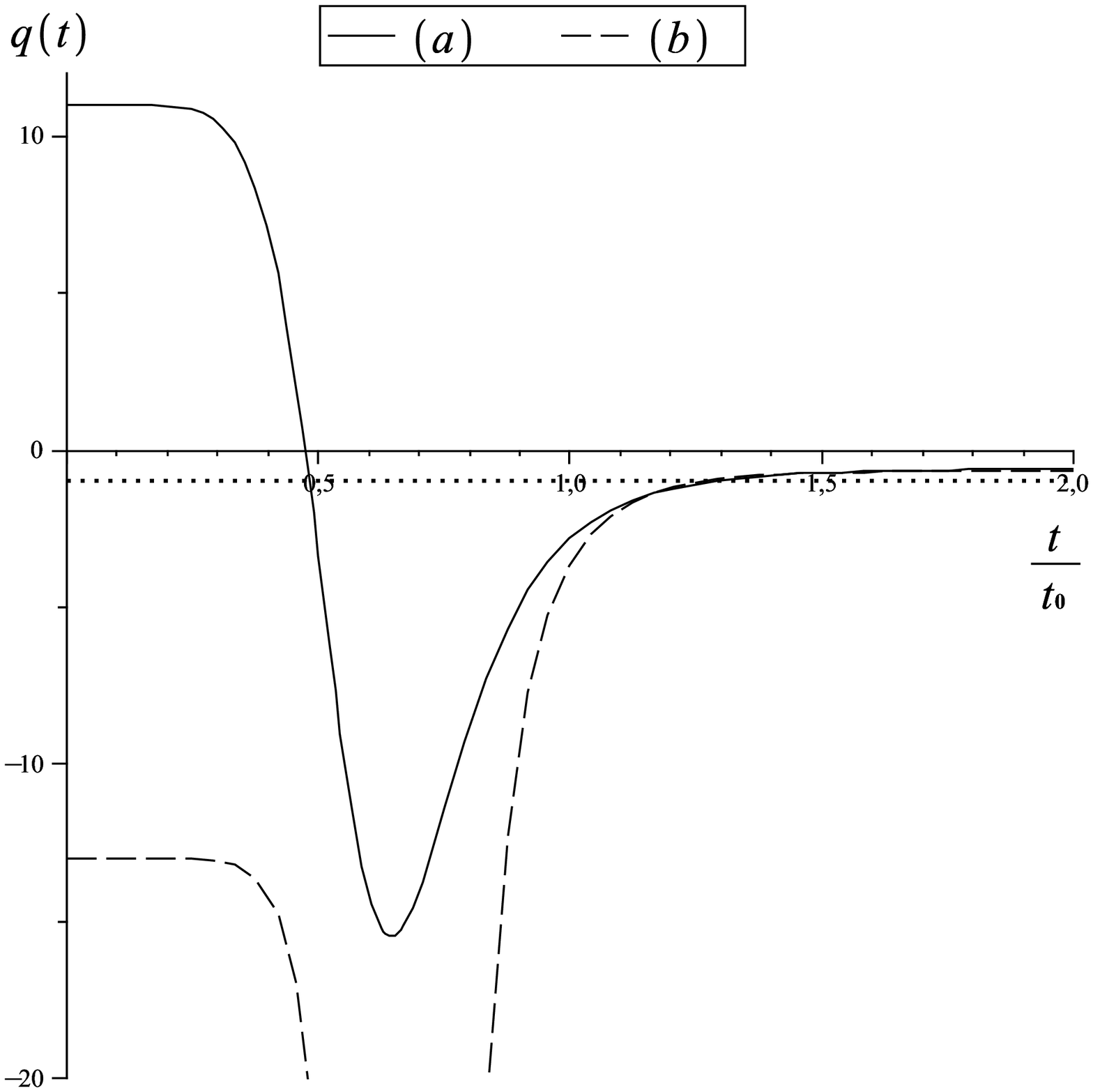} \hfill
\includegraphics[width=0.47\textwidth]{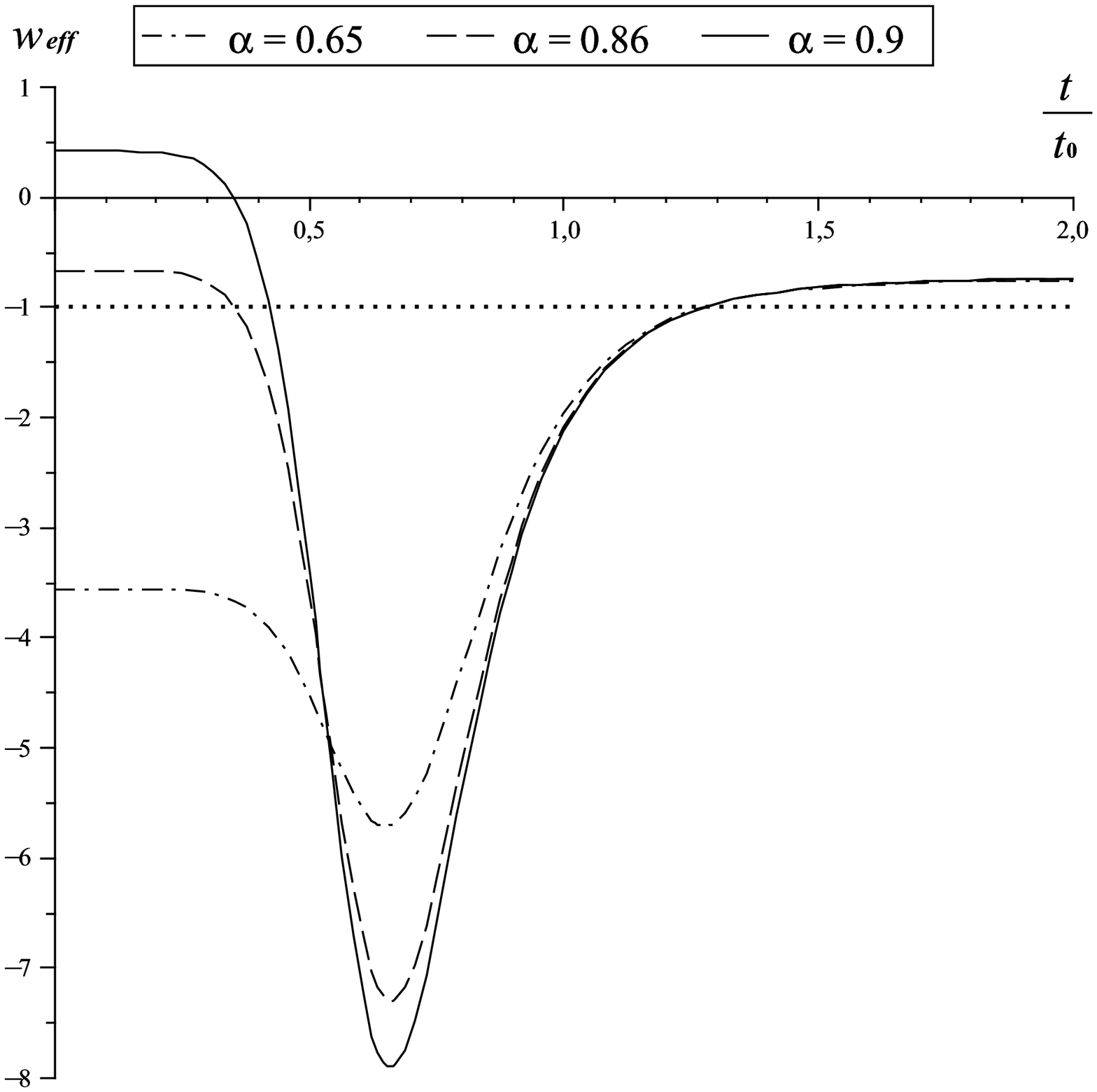}
\parbox[t]{0.47\textwidth}{\caption{The deceleration parameter $q(t)$, corresponding to the cases $(a)$ and $(b)$, represented in Fig. 2.}\label{fig3}} \hfill
\parbox[t]{0.47\textwidth}{\caption{The effective EoS $w_{eff}$, corresponding to the case $(a)$ in Fig. 2 as a function of time for different $\alpha $.}\label{fig4}}
\end{figure}

Note one more feature of the model, associated with a particular type of symmetry of the equation (\ref{18}). It is easy to see that $X(x) = 1/Y (x)$ satisfies equation of the same type as (\ref{18}) with replacement of the coefficients (\ref{19}) according to  $A \to A; \, B \to -B; \, C \leftrightarrow D $ provided $C, D \ne 0$, that is the equation  $A X'(x) + BX(x) + DX^2(x) = C$. Therefore, the solution for  $X(x) = Y^{-1}(x)$ can be represented as follows:
\begin{equation}\label{21}
X(x) = \frac{1}{2D}\left[-B+\sqrt{B^2+4CD}\,\tanh \left(\frac{\sqrt{B^2+4CD}}{2A}(x+x_0)\right)\right],
\end{equation}
where the constant $x_0$ is determined by the condition $\displaystyle \tanh \frac{\sqrt{B^2 +4 CD}}{2A} x_0 = \frac{\sqrt{B^2 +4CD}}{B}$. After that, we can see that the effective equation of state (\ref{8}) can be simply expressed in terms of the function $X(x)$ of (\ref {21}) as follows:
\begin{equation}\label{22}
w_{eff}(x)= -1 +\frac{1}{3}\,\,\frac{2X'+(3-\alpha)X-(1-\alpha)(2-\alpha)X^2}{1+(1-\alpha)X}.
\end{equation}

Returning to the original variables according to (\ref{15}), solution of the equation (\ref{20}) for the Hubble parameter can be written as
\begin{equation}\label{23}
H(t)=\frac{1}{2 C t}\Big(B+2 n\,A\, \tanh [n\,\ln(t/t_0)]\Big),\,\,\,n=\frac{\sqrt{B^2+4CD}}{2A},
\end{equation}
which yields the following expression for the scale factor:
\begin{equation}\label{24}
a(t)=a_0\, t^{B/2C} \Big(\cosh[n\ln(t/t_0)]\Big)^{A/C},
\end{equation}
where $a_0$ is a constant of integration, $n \in \mathbb{R}$.  The dependence $H(t)$ and $a(t)$ on time, represented by the formulas (\ref{23}) and (\ref{24}), is shown in Fig. 2. Graphs of the evolution of the deceleration parameter $q(t)$ and the effective EoS $w_{eff}(t)$ for the obtained solutions are shown in Fig. 3 and Fig. 4, respectively. It may be noted that for certain values of the coefficients in (\ref{18}), i.e. the constants $\alpha$ and $k_i$ in the formulas (\ref{19}), the late cosmic acceleration  and the repeated crossing of the phantom divide $w_{eff} = -1$ are possible. Such a behavior of the universe is supposed by the modern astrophysical observations \cite{Riess}, \cite{Perlmutter}.

\begin{figure}[t]
\centering
\includegraphics[width=0.47\textwidth]{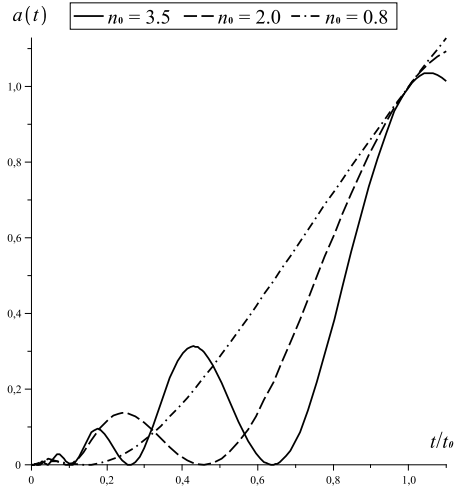}\\
\caption{The scale factor as a function of time for the model $n = i n_0$. Here, $A = 6, \, B = 8, \, C = 3$ and $a_0 = 1$.}
\label{fig5}
\end{figure}

The evolution of the model qualitatively changes in the case $B^2 +4CD < 0$, which is permissible due to the arbitrariness of the constants $k_i$ in the definition (\ref{19}). Denoting in this case a purely imaginary coefficient $n = i n_0$, where $n_0 = \sqrt{|B^2+4CD|} \in \mathbb{R}$, we can write the expression for the scale factor similar to (\ref{24}), but with the substitution $\cosh [n \ln (t/t_0)] \to | \cos [n_0 \ln (t/t_0)] |$. This means that the evolution of the scale factor acquires the cyclic property on a logarithmic time variable $x =\ln(t/t_0)$. Graphs of the function $a(t)$ for various values of $n_0$ are shown in Fig. 5. Note that the evolution of the scale factor of the cyclic nature is previously discussed in the literature in different contexts.

Finally, let us consider some examples of with the phenomenological functions $\Lambda(t)$, having an observational basis and widely discussed in the literature (see, e.g. \cite{Overduin}, \cite{Sahni}). To this end, we rewrite the {\it ansatz} (\ref{17})  in terms of the original variables using the definitions (\ref{15}) in the form
\begin{equation}\label{25}
\dot \Lambda = (1-\alpha)\Big[ k_1\frac{\dot H}{t}+(k_1+k_2)\frac{H}{t^2}+k_3\frac{H^2}{t}+\frac{k_4}{t^3}\Big]
\end{equation}
and consider the following cases.

(1) Let $k_i = 0$, where $i = 1,..,4 $. Then from (\ref{25}), it follows that $\Lambda = constant$. In accordance with the definitions (\ref{19}) and (\ref{23}), we obtain:
$$
A_{1} = 1,\,\,\,B_{1} = 9-2\alpha,\,\,\,C_{1} = 3,\,\,\,n_{1}=\frac{1}{2}\sqrt{(9-4\alpha)^2+24}.
$$

(2) Now let $k_1 = k_2 = k_3 = 0 $ and $ k_4 = -2 \beta/(1 - \alpha)$, where $\beta$ is an arbitrary positive constant. Integrating (\ref{25}), we obtain $\displaystyle \Lambda (t) = \frac{\beta}{t^2}$, where the integration constant is assumed to be zero, and from (\ref{19}) and (\ref{23}) we find the following coefficients:
$$
A_{2} = 1,\,\,\,B_{2} = 9-2\alpha,\,\,\,C_{2} = 3,\,\,\,n_{2}=\frac{1}{2}\sqrt{(9-4\alpha)^2+24\frac{1-\alpha-\beta}{1-\alpha}}.
$$

(3) Finally, let $\displaystyle k_1 = \frac{\beta}{1 - \beta}, \, k_2 =-2k_1, \, k_3 = k_4 = 0$. From equation (\ref{25}) it follows that $\displaystyle \Lambda(t) = \frac{\beta}{t}H(t)$, and the constant coefficients of (\ref{19}) and (\ref{23}) are given by the following expressions:
\begin{eqnarray}
A_{3} &=& \frac{1-\alpha-\beta}{1-\alpha},\,\,\,B_{3} = 9-2\alpha-\frac{2\beta}{1-\alpha},\,\,\,C_{3} = 3,\,\,\,\nonumber \\
n_{3}&=&\frac{1}{2(1-\alpha-\beta) }\sqrt{\Big[(1-\alpha)(9-4\alpha)-2\beta\Big]^2+8(1-\alpha)\Big[3(1-\alpha)-\alpha\beta\Big]}. \nonumber
\end{eqnarray}
It is interesting to note that the parameter $n_{2}$ can become imaginary for the coupling constant $\beta > (1 - \alpha) [(9-4 \alpha)^2 + 24] / 24$, which corresponds to the case illustrated in Fig. 5.

\section{Conclusion}

Thus, in this paper we have obtained some exact solutions of the cosmological model derived from the variational principle for fractional action functional in \cite{Shchigolev}. Exact solutions for the dynamical equations of this model are found either under the assumption of a vacuum-like state of matter that fills the universe, or on the basis of a rather  general {\it ansatz} for the dynamical cosmological term. In any case, the behavior of the model demonstrates a significant difference from the corresponding standard model, which obviously is a consequence of the fractional order of action functional  and, possibly, the fractal nature of space-time (see e.g. \cite{1Calcagni}, \cite {2Calcagni}).

Particularly attractive in our opinion is that the effective EoS which is defined by (\ref{8}), can cross the phantom divide line, as supported by the recent astrophysical observations. Recall that in standard cosmology with the sole source this crossing is not possible. In addition, the model is able to evolve cyclicly, that is attracting the attention of researchers for the recent years (see \cite{Novello} for a review).

Finally, we have to note that our ansatz (\ref{17}) for the cosmological term proposed in the present work,  can not cover all the features of the models in FAC. Therefore, it would be interesting  to continue this research towards a more general ansatz, which could include the greater number of  phenomenological expressions for $\Lambda(t)$.

\end{document}